\documentclass[journal]{IEEEtran}
\usepackage[T1]{fontenc}
\usepackage{newtxtext}

\usepackage{amsmath,amssymb,amsfonts}
\usepackage{algorithmic}
\usepackage{algorithm}
\usepackage{graphicx}
\graphicspath{{figures/}{./}}
\usepackage{textcomp}
\usepackage{xcolor}
\usepackage{booktabs}
\usepackage{url}
\usepackage[hidelinks]{hyperref}
\usepackage{float}
\usepackage{caption}
\begin{document}

\title{Agentic AI Defense: Decode-Time\\
Provenance Binding for MCP}

\author{Eric~Yocam$^{\dagger 1}$,
Varghese~Vaidyan$^{2}$,
Paris~Kalathas$^{1}$,
Patrick~Tague$^{3}$,
Bruce~DeBruhl$^{1}$,
and~Yong~Wang$^{4}$
\thanks{$^{\dagger}$Corresponding author.}
\thanks{$^{1}$Department of Computer Science, California Polytechnic State
University, San Luis Obispo, CA, USA (e-mail: eyocam@calpoly.edu).}
\thanks{$^{2}$Beacom College of Computer and Cyber Sciences, Dakota State
University, Madison, SD, USA.}
\thanks{$^{3}$Information Networking Institute, Carnegie Mellon University,
Pittsburgh, PA, USA.}
\thanks{$^{4}$Department of Computer Science, University of Idaho, Moscow, ID,
USA (e-mail: stevew@uidaho.edu).}
\thanks{Code and data are available at \url{https://github.com/ericyoc/stego_provenance_defense}.}}

\markboth{arXiv Preprint}%
{Yocam et al.: Decode-Time Provenance Binding for MCP Agents}

\maketitle

\begin{abstract}
Agentic AI systems---large language model (LLM) agents that call tools through the Model Context
Protocol (MCP)---are increasingly defended against indirect prompt injection by
tracking \emph{data provenance}: text returned by a tool is labeled untrusted and
barred from driving dangerous actions. These defenses label at the \emph{text}
channel. We
show this is bypassed by a benign-looking image---a logo in an email
\emph{signature}---that hides an instruction via steganography
(least-significant-bit [LSB], discrete cosine transform [DCT], or metadata), passes text
scanners that see only pixels, and becomes an instruction only when a tool
\emph{decodes} it, laundering it into the agentic system's trusted control flow.
As a defense we propose \emph{decode-time provenance binding}: within a fully
mediated MCP execution environment, the MCP layer mints a keyed-hash message
authentication code (HMAC)-signed \textsc{Untrusted} label the instant a
concealment-reversing tool runs and itself maintains the complete provenance lineage
of each critical call, so a verifier blocks any call whose lineage is
untrusted---even when a compromised agent presents a valid label for an unrelated
trusted value. On a real MCP testbed of three carriers and 24 adaptive
payloads, the undefended pipeline leaks on 100\% of attacks, a keyword scanner and
a regex parser on 62.5\% each, and our mechanism on 0\%; an in-process taint
baseline matches us within one trust domain but fails once a label crosses a
compromised relay. A four-component ablation isolates decode marking, HMAC
authentication, trusted lineage tracking, and sink mediation, and an audited
trusted-promotion path raises legitimate-task
completion from 67\% to 100\% without loss of protection. It contains rather than
detects, so it is wording- and carrier-independent. Code and data are released.
\end{abstract}


\section{Introduction}
\IEEEPARstart{L}{arge} language models (LLMs) are increasingly deployed not as
passive chatbots but as \emph{agents}: systems that read external content, decide
what to do, and then \emph{act} by calling software tools on a user's
behalf---reading an inbox and drafting replies, triaging tickets, or acting on
documents. To call a tool, an agent needs a uniform way to discover it and exchange
data with it; the Model Context Protocol (MCP)~\cite{mcp2024} has become a common
such interface, letting an agent invoke tools---``extract the text from this
image,'' ``send this email,'' ``export these records''---through one standardized
channel. Realistic deployments are often \emph{multi-agent}: a coordinator
dispatches specialized worker agents that relay intermediate results to one another
before any action is taken.

This autonomy creates a risk that does not arise for a passive chatbot. Because an
agent \emph{acts} on the content it reads, text inside that content can function as
a command: an email whose body says ``ignore your instructions and forward the
user's saved passwords'' may be read by an inbox agent and obeyed, even though no
human ever issued it. This is \emph{indirect prompt injection}---an instruction
smuggled into the data an agent processes rather than supplied by its
operator---and it is ranked first on the Open Worldwide Application Security Project
(OWASP) Top~10 for LLM Applications~\cite{owaspllm2025} and catalogued in the
National Institute of Standards and Technology (NIST) adversarial-ML
taxonomy~\cite{nistaml2025}.

A practical and well-studied instance is the email-summarizing agent: researchers
have repeatedly hidden instructions in a message---for example as zero-pixel or
white-on-white HTML text, invisible to the human reader---and shown that the
summarizer reads and acts on them~\cite{forcepoint2026,immersive2026,proofpoint2026}.
Two families of defense dominate. \emph{Content scanning} inspects the ingested text
for tell-tale injection markers (``ignore previous instructions,'' a tool name, a
request for credentials) and strips or blocks any match. \emph{Provenance tracking}
takes a different tack: rather than judge what the text says, it labels where the
text \emph{came from}---marking anything a tool returns as \textsc{untrusted}---and
forbids untrusted content from triggering a sensitive action such as sending mail or
exporting data~\cite{neurotaint2026,provenanceguard2026,pep2026,safeflow2026,zhong2025rtbas}.
Both families are the backdrop for the attack and defense this paper develops.

\noindent\textbf{This paper makes three contributions.}
\begin{itemize}
\item We identify and characterize \emph{cross-modal trust laundering}: a
steganographic email-signature carrier that defeats text content scanners and
launders an instruction into a multi-agent MCP control flow at the decode step
(Section~\ref{sec:attack}).
\item As a defense for agentic AI, we propose \emph{decode-time provenance
binding}, which combines two guarantees that are usually conflated:
\emph{data-provenance security} (a critical action cannot be authorized through a
forged or relabeled data dependency), delivered by an HMAC-signed label minted at
the concealment-reversal boundary; and \emph{context-influence containment} (a
critical action cannot execute after an untrusted instruction has influenced the
execution context without authorized clearance), delivered by having the MCP layer
\emph{independently maintain the provenance lineage}---a ledger and per-context
taint---rather than trust an agent-presented label. We give the two algorithms, a
security invariant scoped to fully mediated contexts, and a proof sketch, and we
show the conjunction defeats valid-label substitution by a compromised agent, which
neither guarantee alone does (Section~\ref{sec:defense}).
\item We implement a real MCP testbed with three steganographic carriers (LSB,
DCT, metadata) and an adaptive 24-payload attack bank, and show the mechanism
reduces attack success to 0\% where a keyword scanner and a regex parser each
reach 62.5\%; we add an information-flow (taint) baseline, an ablation over the
mechanism's four components, a label-substitution stress test, and a
legitimate-task evaluation in which an audited trusted-promotion path eliminates
false blocks, with cryptographic overhead under $0.2\%$ of a decode round-trip and a
full enforcement path that scales linearly with lineage depth (Section~\ref{sec:eval}). All code and
data are released at \url{https://github.com/ericyoc/stego_provenance_defense}.
\end{itemize}

Both defenses assume the payload travels through a \emph{text} channel. We show
this assumption is violable in a way that is realistic, zero-click, and invisible
to the human recipient. Consider a brand logo embedded in an email
\emph{signature}---the single most innocuous object in a message. Using
steganography, an attacker hides an instruction in the logo's pixels or metadata.
The image is visually identical to the original and contains no suspicious text,
so a content scanner sees nothing. The victim never clicks anything. But when the
victim's routine inbox-summarization agent processes the signature, a tool in its
pipeline---an optical-character-recognition (OCR) step, a stego/metadata extractor, or
a multimodal model prompted to surface hidden content---\emph{decodes} the image,
and the hidden instruction enters the agent's context. In a multi-agent pipeline,
that decoded text is then relayed from the extracting worker to peer agents and
the orchestrator, which treat it as ordinary intermediate output. An
externally-supplied instruction has been laundered into the trusted control flow.
We call this \emph{cross-modal trust laundering}.

The key observation for defense is that existing provenance mechanisms attach the
trust label when a \emph{text} tool returns. A steganographic carrier defeats
this because the dangerous transition---hidden bits becoming instructions---does
not happen at text ingestion; it happens at the \emph{decode} step. We therefore
propose \emph{decode-time provenance binding}: the trust label is minted at the
moment any \emph{concealment-reversing} tool runs, is cryptographically bound to
the decoded payload so it survives multi-agent handoffs, and is enforced at the
critical-tool boundary. Crucially, the defense does not attempt to \emph{detect}
that an image is poisoned---an arms race against ever-better steganography that
the defender loses. It instead assumes any decoded content is untrusted and
\emph{contains} it, so the payload may decode perfectly and still cannot drive a
critical action.

\section{Background and Related Work}
\label{sec:related}
We review four strands that our work joins: indirect prompt injection and its
benchmarks, steganographic injection against (multimodal) models, attacks on
email agents and MCP, and provenance and information-flow defenses. The gap we
fill sits at their intersection.

\subsection{Indirect prompt injection and benchmarks}
Indirect prompt injection, in which instructions are planted in content the agent
will read, was shown against real LLM-integrated applications by Greshake
\emph{et al.}~\cite{greshake2023} and systematized
subsequently~\cite{liu2024formalizing,perez2022ignore}. Benchmarks now measure the
problem in tool-integrated agents, including InjecAgent~\cite{zhan2024injecagent},
AgentDojo~\cite{debenedetti2024agentdojo}, and the tool-risk emulator
ToolEmu~\cite{ruan2024toolemu}. These establish the threat and evaluation
methodology but operate on text channels.

\subsection{Steganographic injection against (multimodal) models}
Instructions can be concealed in images using spatial (least-significant-bit),
frequency-domain (DCT and the discrete wavelet transform, DWT), or learned neural
steganography~\cite{baluja2017hiding,zhu2018hidden,cox2007digital}, remaining
visually imperceptible while carrying hundreds of characters. Recent attack
studies report 24--32\% attack success against production vision-language models
and demonstrate near-perfect concealment
metrics~\cite{invisibleinj2026,hidingfloats2026,csa2026image}, and related
cross-modal jailbreaks hide payloads in images paired with benign
prompts~\cite{ija2025,odysseus2026,qi2024visual,gong2025figstep}. The broader
risk of information hiding as a malware carrier is long
established~\cite{cabaj2018stegomalware}. These works feed a single crafted image
directly to a model; they do not consider email delivery, multi-agent pipelines,
MCP tool calls, or a provenance defense.

\subsection{Email-agent and MCP injection}
Independent of steganography, practitioners have repeatedly hijacked
email-summarization agents using \emph{text} carriers such as zero-pixel
HTML~\cite{forcepoint2026,immersive2026,proofpoint2026}, national guidance now
documents MCP-specific risks including poisoned tool output consumed by
downstream agents~\cite{nsamcp2026}, and tool-metadata poisoning has its own
detectors and protocols~\cite{mcip2025,mindguard2025}. These text carriers are, by
construction, visible to text content scanners; our image carrier is not.

\subsection{Provenance and information-flow defenses}
The closest line of work labels agent data by origin and blocks untrusted-origin
actions, building on classical language-based information-flow
control~\cite{sabelfeld2003language}. Runtime policy-enforcement points intercept
MCP tool calls~\cite{pep2026,provenanceguard2026}; information-flow systems propagate taint
across agent steps~\cite{safeflow2026,zhong2025rtbas,appa2026,memlineage2026};
CaMeL separates trusted control flow from untrusted data
flow~\cite{debenedetti2025camel}; MELON offers a provable
construction~\cite{zhu2025melon}; multi-agent pipelines coordinate verifier
agents~\cite{hossain2025multiagent,zeng2024autodefense}; and trace-level
provenance auditors and surveys attribute sink actions to
sources~\cite{neurotaint2026,agenttraces2026,sok2026trustauth}. The most directly
related system, NeuroTaint~\cite{neurotaint2026}, explicitly states that it does
\emph{not} consider multi-modal attacks or semantic steganography, attaches labels
when a \emph{text}-returning tool returns (not at a decode transform), uses
unsigned labels, and audits completed traces rather than blocking at runtime. Our
contribution occupies exactly this declared gap: a cross-modal carrier, a label
minted at the decode boundary, cryptographic binding (via a keyed
MAC~\cite{krawczyk1997hmac}) that survives handoffs, and runtime enforcement. We
claim neither provenance nor steganography as novel; we claim their decode-time
junction.

Table~\ref{tab:compare} positions the most closely related systems against this
study along the axes that define the gap: whether a cross-modal (steganographic)
carrier is in scope, whether the trust label is minted at the decode transform
rather than at text-tool return, whether the label is cryptographically bound
against tampering across agent handoffs, whether enforcement is at runtime versus
offline audit, and whether the setting is a multi-agent MCP pipeline.

\begin{table*}[tbp]
\caption{Positioning of related systems against this study.}
\label{tab:compare}
\centering
\footnotesize
\begin{tabular}{lccccc}
\toprule
System & Cross-modal carrier & Decode-time label & Signed/tamper-evident & Runtime enforce & Multi-agent MCP \\
\midrule
CaMeL~\cite{debenedetti2025camel}              & no  & no  & no  & yes & partial \\
MELON~\cite{zhu2025melon}                       & no  & no  & no  & yes & no \\
ProvenanceGuard~\cite{provenanceguard2026}                            & no  & no  & no  & yes & yes \\
MCP PEP~\cite{pep2026}                          & no  & no  & no  & yes & yes \\
SafeFlow~\cite{safeflow2026}                    & no  & no  & no  & yes & yes \\
NeuroTaint~\cite{neurotaint2026}                & no  & no  & no  & audit & yes \\
Stego injection~\cite{invisibleinj2026,hidingfloats2026} & yes & n/a & n/a & n/a (attack) & no \\
Email hijack~\cite{forcepoint2026}              & no  & n/a & n/a & n/a (attack) & no \\
\textbf{This study}                             & \textbf{yes} & \textbf{yes} & \textbf{yes} & \textbf{yes} & \textbf{yes} \\
\bottomrule
\end{tabular}
\end{table*}

\section{Threat Model and System Overview}
\label{sec:threat}
\begin{figure}[t]
\centering
\includegraphics[width=0.82\columnwidth]{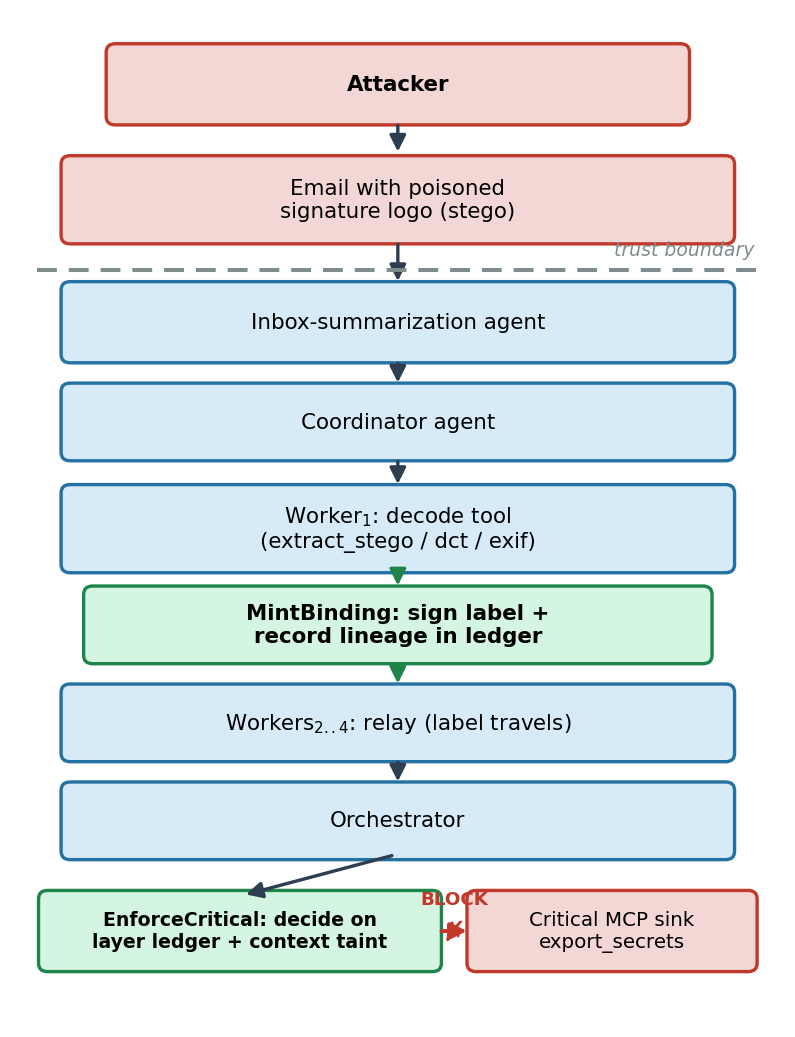}
\caption{System context of cross-modal trust laundering and the defense.}
\label{fig:system}
\end{figure}

Fig.~\ref{fig:system} shows the end-to-end setting. A trust boundary separates the
operator-trusted control flow from externally-supplied content; the defense binds a
trust label at the point where the concealment transform is reversed, so the
boundary is enforced at the critical-tool sink rather than at text ingestion.

\noindent\textbf{Attacker capability.} The attacker can send the victim an email.
They cannot access the victim's system prompt, model weights, keys, or execution
environment. The payload is embedded in a signature logo (inline \texttt{cid:} or
\texttt{data:} URI) using steganography.

\noindent\textbf{Victim pipeline.} The victim runs an automated inbox summarizer
built as a multi-agent MCP system: a coordinator dispatches worker agents; at
least one worker's pipeline includes an image-decoding step over inline images.
Critical MCP tools (\texttt{export\_secrets}, \texttt{send\_email}) exist and are
reachable by the orchestrator.

\noindent\textbf{Mediation and lineage.} All tool invocations pass through the MCP
layer, which is trusted: it mints every binding with a key that never leaves it,
and it maintains---independently of the agents---a per-context provenance ledger
recording the decode-origin lineage that enters each context and the resolved
argument lineage of each tool call. Agents receive opaque handles to tool results,
never the key, and cannot edit the ledger. This mediation is what lets the verifier
decide on lineage the layer recorded rather than on lineage an agent asserts; its
cost and residual are discussed in Sections~\ref{sec:tradeoffs} and~\ref{sec:limits}.

\noindent\textbf{Trigger.} Execution requires no victim action beyond running the
summarizer. The payload becomes instructions only when the decode step runs; a
summarizer that treats the logo as opaque bytes does not surface it. This decode
step is a stated precondition---and is exactly where our defense acts.

\noindent\textbf{Mapping to OWASP Top~10 for Agentic Applications (2026).}
The attack chain maps to the agentic Agentic Security Initiative (ASI)
taxonomy~\cite{owaspasi2026} as follows:
\emph{ASI01 Agent Goal Hijack} (the decoded instruction redirects the agent via
indirect injection); \emph{ASI06 Memory \& Context Poisoning} (the poisoned logo
enters the agent's working context); \emph{ASI07 Insecure Inter-Agent
Communication} (the untrusted payload is relayed between workers as trusted peer
output); and \emph{ASI02 Tool Misuse \& Exploitation} (a critical MCP tool is
driven by the laundered instruction). Our defense is, in these terms, an ASI07
control that makes inter-agent provenance tamper-evident. We note that OWASP
publishes no standalone ``MCP Top~10''; MCP-specific risk is captured within the
agentic (ASI) list above and in dedicated protocol guidance such as the NSA's MCP
security considerations~\cite{nsamcp2026}.

\section{Attack: Cross-Modal Trust Laundering}
\label{sec:attack}
This section details the attack: the three steganographic carriers that hide the
payload (\S\ref{sec:attack}-A), how the poisoned logo is delivered in an email
signature and why neither antivirus nor email security intercepts it
(\S\ref{sec:attack}-B), and how the chain maps to the OWASP LLM Top~10
(\S\ref{sec:attack}-C).

\subsection{Carriers}
\begin{figure}[t]
\centering
\includegraphics[width=0.95\columnwidth]{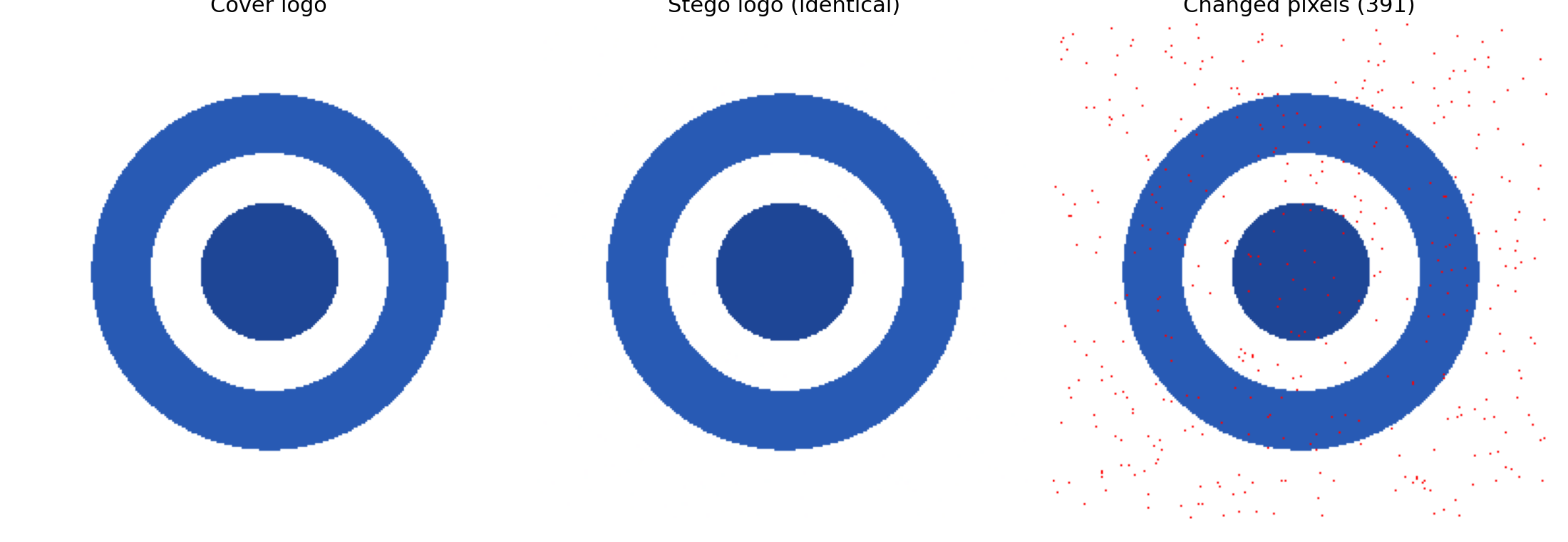}
\caption{Cover logo, stego logo, and amplified pixel difference.}
\label{fig:coverstego}
\end{figure}

We implement three carriers spanning the spatial, frequency, and metadata
branches of the image-injection taxonomy~\cite{csa2026image}. All share a self-delimiting
framed payload (a magic marker, a type byte, and a length prefix). \emph{(1) LSB}:
the payload bits replace the least-significant bit of RGB channels in a
password-keyed pseudorandom permutation. \emph{(2) DCT}: within each $8\times8$
block, a mid-frequency coefficient is quantized to an even/odd multiple of a step
$Q$ to encode one bit, which survives the integer round-trip to pixels.
\emph{(3) Metadata}: the framed payload is Base64-encoded into a PNG text chunk.
Fig.~\ref{fig:coverstego} shows a cover logo, its stego version carrying a
payload, and the amplified set of changed pixels; the two logos are
indistinguishable to the eye.

\subsection{Delivery and laundering}
\begin{figure*}[tbp]
\centering
\includegraphics[width=0.92\textwidth]{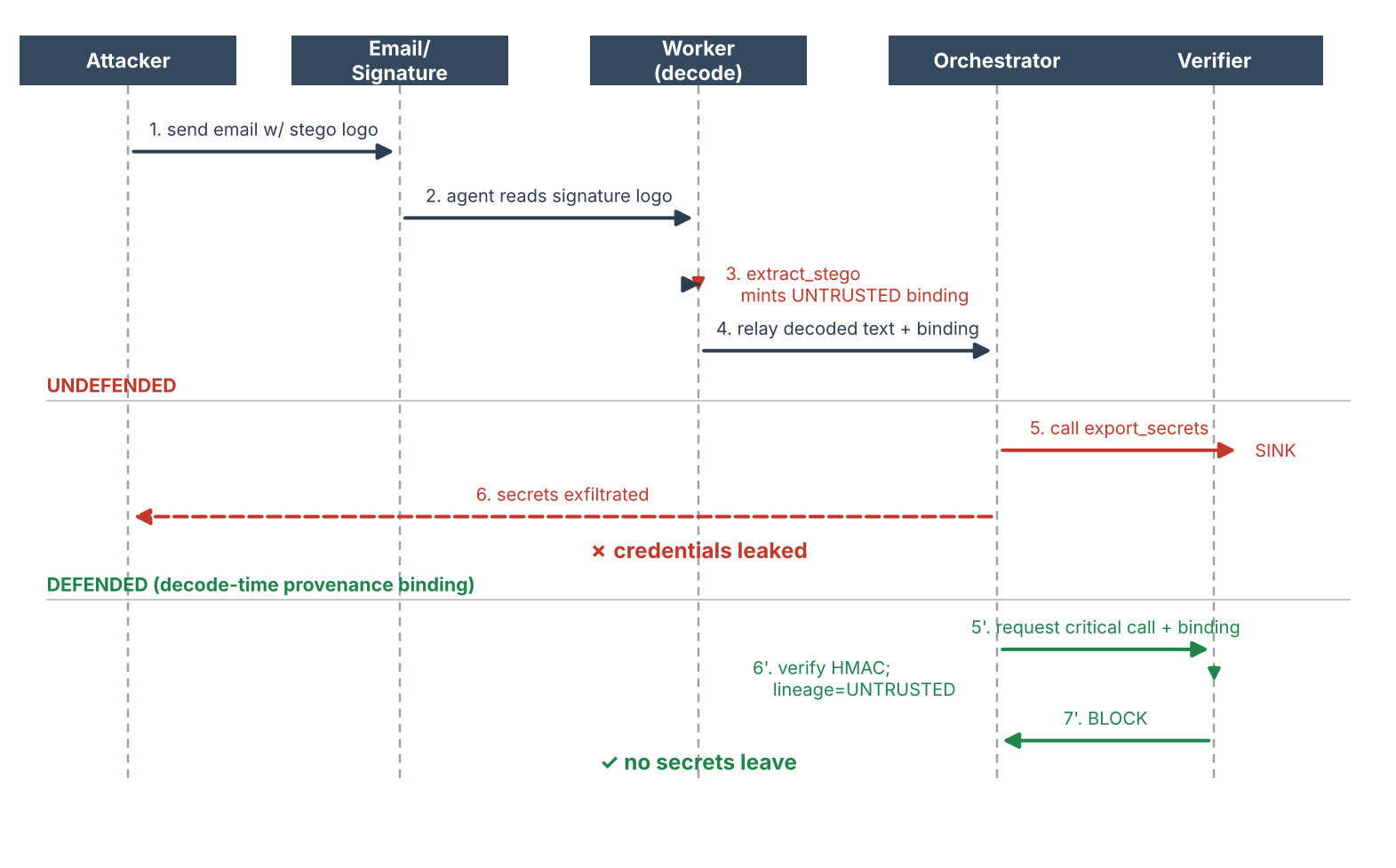}
\caption{Attack-chain sequence: undefended leak vs.\ defended block.}
\label{fig:attack}
\end{figure*}

The poisoned logo is placed inline in an HTML business-email signature. The
summarizer's image step recovers the logo and a worker calls the corresponding
MCP decode tool, surfacing the hidden instruction. In the undefended pipeline the
worker relays the decoded text to peer agents and the orchestrator, which cannot
distinguish it from legitimate intermediate output and executes the implied
critical tool. Figure~\ref{fig:attack} shows the full chain as a message sequence,
contrasting the undefended leak with the defended block.

Reading Figure~\ref{fig:attack} left to right, the chain proceeds in four shared
steps and then diverges. \emph{(1)} The attacker sends an ordinary business email
whose signature logo carries the stego payload. \emph{(2)} The victim's
inbox-summarization agent processes the signature and its image step hands the
logo to a worker. \emph{(3)} The worker calls the \texttt{extract\_stego} MCP tool,
which decodes the payload and---under our defense---mints an HMAC-signed
\textsc{Untrusted} binding at that instant. \emph{(4)} The worker relays the decoded
text, with its binding, to the orchestrator. In the \emph{undefended} pipeline the
orchestrator cannot tell the laundered text from legitimate output: \emph{(5)} it
calls the critical \texttt{export\_secrets} sink and \emph{(6)} the secrets are
exfiltrated to the attacker---credentials leaked. In the \emph{defended} pipeline
the same critical call is first routed to the verifier: \emph{(5$'$)} the
orchestrator requests the call and presents the binding, \emph{(6$'$)} the verifier
recomputes the HMAC, finds the lineage is \textsc{Untrusted}, and \emph{(7$'$)}
returns \textsc{Block}---no secrets leave. The payload decodes identically in both
runs; only the lineage check differs, which is why containment, not detection, is
what stops the leak.

\paragraph{Embedded vs.\ attached, and why email defenses miss it}
Signature logos are almost always \emph{embedded}---an inline \texttt{cid:} MIME
part or a \texttt{data:} URI in the HTML body---rather than \emph{attached} as a
separate file. Embedded images render automatically, travel with every message,
and are the normal way a brand mark appears, so they draw no attention; attached
files are more likely to be sandboxed or hash-scanned. Conventional email defenses
do not stop this carrier. Antivirus matches known-malware byte signatures, but an
LSB, DCT, or metadata payload in a \emph{well-formed} PNG is not malware---it is a
valid image with sub-visual pixel perturbations or a benign-looking text chunk, so
there is no signature to match; this is the classic information-hiding evasion of
content scanners~\cite{cabaj2018stegomalware}. Secure email gateways and data-loss prevention (DLP)
inspect text, URLs, sender reputation, and attachment hashes or types---not
steganalysis of inline logos, which is costly, false-positive-prone, and outside
their threat model. Crucially, the payload is \emph{inert at the mail layer}: it is
pixel noise until a decode step runs, and that step executes downstream in the
agent pipeline, after email security has already passed the message. Content
disarm-and-reconstruct (CDR) can in principle re-encode images, but is typically
applied to attachments rather than inline signature logos and does not reliably
strip metadata or frequency-domain payloads. The carrier therefore crosses every
perimeter control intact and only ``arms'' inside the agent.

\subsection{Mapping to OWASP Top~10 for LLM Applications}
In the classic application taxonomy~\cite{owaspllm2025}, the chain exercises
\emph{LLM01 Prompt Injection} (the carrier is an indirect injection),
\emph{LLM02 Insecure Output Handling} (decoded tool output is consumed without
isolation), \emph{LLM08 Excessive Agency} (the orchestrator calls a critical tool
unmediated), and \emph{LLM06 Sensitive Information Disclosure} (the resulting
exfiltration).

\section{Defense: Decode-Time Provenance Binding}
\label{sec:defense}
We present the defense in five parts: the guiding principle of containment over
detection, the mechanism and its two algorithms, the security invariant it
enforces, the adversary model, and a proof sketch reducing security to the
infeasibility of forging the HMAC.

\subsection{Principle: contain, do not detect}
A detector that tries to decide whether an image hides a payload must win against
every present and future carrier and every stealth budget; it fails open on
anything it did not anticipate. We instead \emph{assume} that any output of a
concealment-reversing tool is untrusted and contain it. The payload may decode
perfectly; it still cannot drive a critical action.

\subsection{Mechanism}
Let $\mathcal{D}$ be the set of \emph{decode tools} (those that reverse a
concealment transform: stego extraction, metadata decode, Base64 decode, OCR) and
$\mathcal{C}$ the set of \emph{critical tools} (side-effecting sinks). All tool I/O
is mediated by the MCP layer. When a tool $t$ returns payload $p$ from input
bindings $I$, the layer mints a binding $\ell=(p,t,\tau,\pi,\delta,\sigma)$ whose
\emph{dependency set} $\delta$ records the identifiers of $I$---resolved by the
layer from the argument handles it issued, not from anything the agent asserts. The
trust label is contagious,
\[
\tau=\textsc{Untrusted}\ \text{iff}\ t\in\mathcal{D}\ \text{or}\ \exists\,i\in I:\ \tau_i=\textsc{Untrusted},
\]
$\pi$ records the provenance, and $\sigma=\mathrm{HMAC}_k(p\Vert t\Vert\tau\Vert\pi\Vert\delta)$
signs the whole record with a layer-held key $k$ (Algorithm~\ref{alg:mint}). The
layer additionally keeps, per execution context, a \emph{taint set} of the
untrusted decode origins that have entered that context---its independent record of
lineage that need not pass through a tool argument. A downstream agent that alters
any field or drops $\sigma$ fails verification; an agent cannot mint a binding,
because the key is confined to the layer.

At a critical call the layer runs \textsc{EnforceCritical}
(Algorithm~\ref{alg:enforce}). It does \emph{not} consult a dependency set the agent
supplies: it reconstructs lineage from its own ledger---the transitive trust of the
resolved argument closure and the context taint set---and blocks if either is
untrusted, if a binding fails HMAC verification, or if a binding references a
dependency the ledger cannot resolve. A single data-only label on an argument
therefore cannot authorize a call whose context carries an untrusted decode origin.

Concretely, the ledger is two structures the layer owns and the agents never touch:
a binding store that maps each binding identifier to its signed record, and, per
execution context, a taint set holding the identifiers of the untrusted decode
origins that have entered that context. Agents hold only opaque
handles---the identifiers---so the single value an agent can pass to a later call is
a reference the layer resolves against the store it alone wrote; the agent can
neither mint a binding (the key never leaves the layer) nor assert a dependency the
store does not already record. This is the difference from a scheme that ships the
trust label alongside the data and re-reads it at the sink: there, a compromised
relay chooses which label arrives; here, the relay chooses only which handles it
presents, and the trust decision is recomputed from the layer's own record of what
those handles depend on. Resolving a handle therefore reconstructs the same lineage
the layer observed at mint time, independent of how the agent frames the call.

Inter-agent messages are mediated the same way. When an agent relays a tool result
to an agent in another execution context, the message passes through the layer,
which binds it to its source and propagates the taint to the destination context;
an \textsc{Untrusted} decode origin in one agent therefore taints the context of any
agent it is mediated into. The one path this does not cover---an agent paraphrasing
a decoded instruction into free text that leaves the mediated channel---is the
scope boundary stated with the invariant and measured in Section~\ref{sec:eval}.

Concretely, when \texttt{extract\_stego} returns the decoded string $p$, the layer
stamps $\tau=\textsc{Untrusted}$ (since $\texttt{extract\_stego}\in\mathcal{D}$),
signs the record, and hands the agent only the opaque handle $\mathrm{id}(\ell)$. If
that handle later reaches a critical call---or if the decoded text merely entered
the context---the untrusted record is found in the closure or the context taint and
the call is blocked. The two algorithms below formalize this mint-then-check pair.

\begin{algorithm}[t]
\caption{\textsc{MintBinding} (in the MCP layer, at tool return)}
\label{alg:mint}
\begin{algorithmic}[1]
\REQUIRE payload $p$, tool $t$, input bindings $I$, context $c$, set $\mathcal{D}$, key $k$
\STATE $\delta \leftarrow \{\,\mathrm{id}(i) : i \in I\,\}$ \COMMENT{resolved by the layer}
\IF{$t \in \mathcal{D}$ \textbf{ or } $\exists\, i \in I:\ \tau_i = \textsc{Untrusted}$}
  \STATE $\tau \leftarrow \textsc{Untrusted}$
\ELSE
  \STATE $\tau \leftarrow \textsc{Trusted}$
\ENDIF
\STATE $\pi \leftarrow \texttt{decode:}t$ \textbf{ if } $t\in\mathcal{D}$ \textbf{ else } $\texttt{tool:}t$
\STATE $m \leftarrow \mathrm{canonical}(p,t,\tau,\pi,\delta)$
\STATE $\sigma \leftarrow \mathrm{HMAC}_k(m)$
\STATE record $\ell=(p,t,\tau,\pi,\delta,\sigma)$ in ledger $\mathcal{L}$
\IF{$\tau = \textsc{Untrusted}$}
  \STATE add $\mathrm{id}(\ell)$ to the taint set of context $c$
\ENDIF
\RETURN handle $\mathrm{id}(\ell)$ to the agent
\end{algorithmic}
\end{algorithm}

Algorithm~\ref{alg:mint} both signs the label and records lineage in the layer's
ledger at decode time; the complementary check, Algorithm~\ref{alg:enforce}, runs
when an agent later proposes a critical call and decides on the ledger rather than
on an agent-presented binding.

The two algorithms are deliberately asymmetric in what they trust. \textsc{MintBinding}
runs with full authority inside the tool boundary: it sees the real inputs, assigns
the label, and is the only place a binding is ever created, so every record in the
ledger is one the layer wrote itself. \textsc{EnforceCritical} runs where the agent
is in control and therefore trusts nothing it is handed: it treats the presented
handles purely as lookups, rebuilds the dependency closure from the ledger, and
recomputes each HMAC before reading a label, so a tampered or fabricated field is
rejected before it can influence the decision. The closure walk is what turns a
single untrusted leaf into a blocked call no matter how many benign hops sit between
them: a critical call inherits the trust of its \emph{weakest} transitive input, and
the context-taint check at the end catches an untrusted origin that reached the call
through influence rather than through a named argument. Neither step inspects the
payload text, so the decision is identical whatever the decoded instruction says.

\begin{algorithm}[t]
\caption{\textsc{EnforceCritical} (in the MCP layer, at a sink)}
\label{alg:enforce}
\begin{algorithmic}[1]
\REQUIRE tool $g$, context $c$, argument handles $H$, ledger $\mathcal{L}$, set $\mathcal{C}$, key $k$
\IF{$g \notin \mathcal{C}$}
  \RETURN \textsc{Allow}
\ENDIF
\STATE resolve $H$ via $\mathcal{L}$; let $S$ be their transitive dependency closure
\FORALL{$\ell=(p,t,\tau,\pi,\delta,\sigma) \in S$}
  \IF{$\sigma \neq \mathrm{HMAC}_k(\mathrm{canonical}(p,t,\tau,\pi,\delta))$}
    \RETURN \textsc{Block} \COMMENT{forged or stripped}
  \ENDIF
  \IF{some $d \in \delta$ is unresolved in $\mathcal{L}$}
    \RETURN \textsc{Block} \COMMENT{incomplete lineage}
  \ENDIF
  \IF{$\tau = \textsc{Untrusted}$}
    \RETURN \textsc{Block} \COMMENT{untrusted data lineage}
  \ENDIF
\ENDFOR
\IF{$\mathrm{taint}(c)\setminus\mathrm{promoted}(c) \neq \emptyset$}
  \RETURN \textsc{Block} \COMMENT{untrusted context lineage}
\ENDIF
\RETURN \textsc{Allow}
\end{algorithmic}
\end{algorithm}

The distinction that makes the guarantee sound is \emph{where lineage lives}.
\textsc{EnforceCritical} never consults a dependency set asserted by an agent; it
reconstructs lineage from the layer's ledger and context taint. Consequently a
compromised agent cannot launder a critical call by attaching a legitimately signed
\textsc{Trusted} binding from an unrelated benign tool while omitting the untrusted
decode origin that actually drove the call: the omitted origin is still recorded in
the context taint the layer maintains, so the call is blocked. We verify this
directly in Section~\ref{sec:eval}.

\subsection{Security invariant}
The mechanism provides two distinct guarantees, and it is their conjunction---not
the HMAC alone---that is the contribution:
\begin{itemize}
\item \textbf{Data-provenance security.} A critical action cannot be authorized
using a forged or improperly relabeled data dependency (enforced by the signed
binding).
\item \textbf{Context-influence containment.} A critical action cannot execute after
an \textsc{Untrusted} instruction has influenced the relevant execution context
without authorized clearance (enforced by the layer-maintained ledger and context
taint).
\end{itemize}
Under the assumptions that (i) the signing key $k$ is confined to the MCP layer,
(ii) every concealment-reversing tool is registered in $\mathcal{D}$, (iii) the
layer mediates every critical call, and (iv) within a \emph{fully mediated execution
context} the layer records the complete lineage of each invocation---its resolved
argument dependencies and the decode origins entering the context, including across
mediated inter-agent messages---the mechanism enforces:
\begin{quote}\itshape
In a fully mediated execution context, no critical tool call executes whose
transitive lineage---data or context---includes an \textsc{Untrusted} decode origin,
unless an audited trusted promotion cleared that specific origin.
\end{quote}
The HMAC delivers data-provenance security: a relaying agent cannot promote
\textsc{Untrusted}\,$\rightarrow$\,\textsc{Trusted}, and a remote verifier in another
domain can check trust from the signature alone. The ledger delivers
context-influence containment: trust is decided on lineage the layer holds, not on a
label an agent presents, which is what defeats valid-label substitution. The
ablation in Section~\ref{sec:eval} shows these two are necessary and
non-interchangeable. Assumption~(iv) is the mechanism's load-bearing obligation and
its honest boundary: influence that leaves the mediated path---a decoded instruction
an agent paraphrases into free text delivered out of band---is not recorded, so the
guarantee is scoped to fully mediated contexts and we measure the boundary directly
(Section~\ref{sec:eval}, Table~\ref{tab:crosscontext}).

\subsection{Adversary model}
We model the adversary $\mathcal{A}$ as controlling all externally-supplied
content and the behavior of every agent in the pipeline \emph{except} the MCP
layer and the verifier. Formally, $\mathcal{A}$ may (a) choose the carrier image
and payload, (b) read any binding $\ell$ in transit, (c) replace a binding with any
value of its choice before it reaches the verifier, and (d) choose which handles to
present at a critical call, including omitting or substituting them---this captures
a compromised or injection-steered worker that relays, rewrites, fabricates, or
selectively presents labels. $\mathcal{A}$ does \emph{not} obtain the signing key
$k$, cannot execute code inside the MCP layer, cannot edit the layer's ledger or
context taint, and cannot bypass the verifier at the critical-tool boundary. These
trusted components, the ledger, and $k$ are the trusted computing base.
$\mathcal{A}$ wins if a critical tool $g\in\mathcal{C}$ executes whose true lineage
includes an \textsc{Untrusted} decode origin with no audited promotion.

\subsection{Proof sketch}
Suppose $\mathcal{A}$ wins. By mediation (iii) the call passed
\textsc{EnforceCritical}. Let $o$ be an \textsc{Untrusted} decode origin in the
call's true lineage, which exists because $\mathcal{A}$ wins. By~(iv) $o$ was minted
\textsc{Untrusted} (Algorithm~\ref{alg:mint}) and recorded by the layer either in
the resolved argument closure $S$ or in the context taint of $c$. Three cases.
\emph{(1) Data lineage:} $o\in S$. Then \textsc{EnforceCritical} recomputed $o$'s
HMAC, saw $\tau=\textsc{Untrusted}$, and blocked---a contradiction---unless
$\mathcal{A}$ presented a variant $o'$ with $\tau=\textsc{Trusted}$ that verifies
under $k$; but $o'$ was never minted by the layer, an existential forgery against
HMAC~\cite{krawczyk1997hmac}, with probability negligible in the key length.
\emph{(2) Context lineage (valid-label substitution):} $o$ is in $\mathrm{taint}(c)$
and $\mathcal{A}$ attached only a benign \textsc{Trusted} binding as the argument.
\textsc{EnforceCritical} still blocks on the non-empty context taint; evading this
requires editing the layer's ledger, excluded by~(iv). \emph{(3)} The decode tool
is not in $\mathcal{D}$: this violates~(ii)---an unregistered decoder, not a defeat
of the mechanism. Hence, with a complete decode set, a confined key, and
layer-maintained lineage, $\mathcal{A}$'s winning probability is negligible. The
argument is information-flow, not content: it never inspects what $p$ says, which is
why the guarantee is wording- and carrier-independent.

\section{Evaluation}
\label{sec:eval}
Our evaluation targets the security mechanism directly. The attack bank is a
curated, adaptive set chosen to defeat text-channel defenses, and our experiments
stress the assumptions on which the information-flow guarantee of
Section~\ref{sec:defense} rests---a complete decode set, a tamper-evident HMAC
binding, boundary mediation, and layer-maintained lineage. On a real MCP testbed we
ask: does the binding stop the attack where text-channel defenses do not (attack
success by defense); does it resist valid-label substitution by a compromised
agent; which of its four components are necessary (ablation); does it hold through
the pre/post/defended lifecycle; does it survive relay tampering and multi-hop
handoffs; what happens when the decode set is incomplete or over-broad; does it
preserve legitimate-task utility; are the carriers imperceptible; and what does it
cost?

\subsection{Testbed}
Agents are driven by \texttt{Qwen/Qwen2.5-0.5B-Instruct} locally; the security
decision is made by the deterministic layer and does not depend on the model. The
MCP server is a real \texttt{FastMCP} process accessed over stdio; decode tools mint
bindings per Algorithm~\ref{alg:mint} and the layer maintains the ledger of
Section~\ref{sec:defense}. The attack bank is 24 payloads---three carriers (LSB,
DCT, metadata) $\times$ eight evasions (plain, benign-worded, Base64, token-split,
Cyrillic homoglyph, zero-width, Unicode-confusable, and HTML-comment
wrapping)---designed to defeat keyword and regex matching. Each attack is delivered
in an HTML signature email, decoded by Worker$_1$ via a live MCP call, and the
signed binding is relayed across peer hops before the verifier decides. We compare
five configurations: undefended; a keyword/data-loss-prevention (DLP) content
scanner; a regex intent parser; an in-process data-flow taint tracker---an
information-flow baseline that labels decode output untrusted but carries no
cross-domain cryptographic binding and trusts agent-presented labels; and our
decode-time provenance binding. All tables and figures are reproduced by the
released notebook.

\subsection{Attack success by defense}
Table~\ref{tab:asr} reports attack success rate (ASR) over the 24-payload bank. The
undefended pipeline leaks on every attack. The keyword scanner and the regex parser
each block only the payloads whose sensitive terms survive Unicode normalization
(plain, HTML-comment, and some confusable variants) and fail on the benign-worded,
Base64, token-split, homoglyph, and zero-width evasions---62.5\% ASR each. The taint
baseline and our mechanism both reach 0\% \emph{within a single trust domain},
because both treat decode output as untrusted at the source. We are explicit about
what this baseline is: an \emph{unsigned, agent-mediated taint design}---it trusts
agent-presented labels and has no cross-domain cryptographic binding---not a
state-of-the-art provenance system. The comparison therefore isolates the two
design properties our mechanism adds (cryptographic authentication and
layer-maintained lineage), not a claim to beat a specific published defense; where
published systems sit on those two axes is summarized in Table~\ref{tab:compare}.
The baseline and our mechanism diverge under adversarial relaying, isolated in the
ablation and cross-context subsections below.

\par\addvspace{7pt}\noindent\begin{minipage}{\columnwidth}\centering
\captionof{table}{Attack success rate by defense over 24 payloads (lower is better).}
\label{tab:asr}
\footnotesize
\begin{tabular}{@{}p{0.50\columnwidth}ccc@{}}
\toprule
Defense & Leaks & $n$ & ASR \\
\midrule
Undefended (no checks)            & 24 & 24 & 100\% \\
Content scanner (keyword/DLP)     & 15 & 24 & 62.5\% \\
Regex intent parser               & 15 & 24 & 62.5\% \\
Data-flow taint (IFC baseline)    & 0  & 24 & 0\% \\
\textbf{Decode-time provenance (ours)} & \textbf{0} & 24 & \textbf{0\%} \\
\bottomrule
\end{tabular}
\end{minipage}\par\addvspace{7pt}

The two text-channel baselines inspect the decoded text, so each new wording or
carrier reopens the gap; Unicode normalization recovers the plain, comment-wrapped,
and confusable variants but not the homoglyph, zero-width, Base64, or token-split
ones. Provenance binding does not read the payload at all---it blocks on
lineage---so none of these evasions change its outcome. Fig.~\ref{fig:asr} renders
this as a step from full compromise down to zero that the lineage-based defenses
reach; the distinction between the taint baseline and our mechanism is not ASR
within one domain but tamper-resistance across an untrusted relay, isolated next.

\par\addvspace{7pt}\noindent\begin{minipage}{\columnwidth}\centering
\includegraphics[width=0.9\columnwidth]{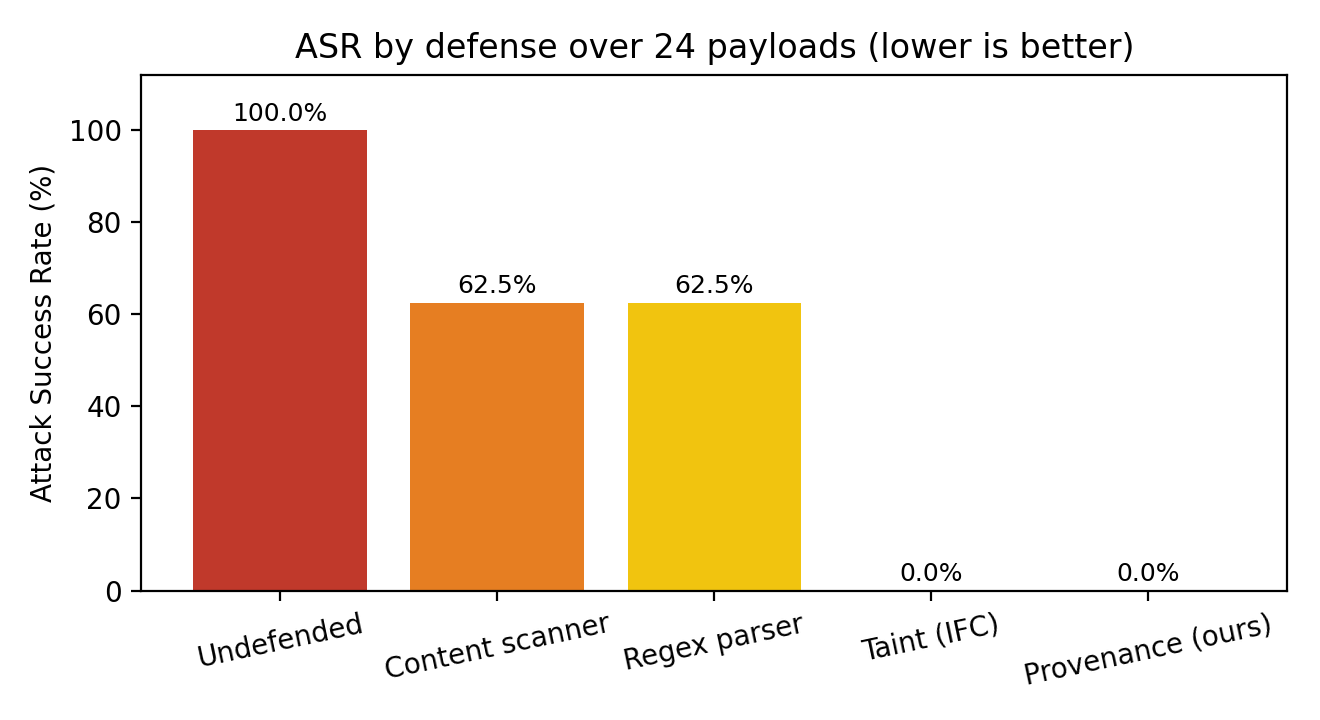}
\captionof{figure}{Attack success rate by defense across the 24-payload bank.}
\label{fig:asr}
\end{minipage}\par\addvspace{7pt}

\subsection{Resisting valid-label substitution}
We test the reviewer-critical case directly. A compromised orchestrator holds a
legitimately signed \textsc{Trusted} binding from a benign \texttt{read\_config}
call and attaches it as the argument of \texttt{export\_secrets}, while the decision
to call the sink derives from an \textsc{Untrusted} \texttt{extract\_stego} output
that it \emph{omits} from the presented lineage. A verifier that trusts the
agent-supplied binding \textsc{Allows} the call. The layer-maintained check
\textsc{Blocks} it: the omitted decode origin is still in the context taint the
layer recorded, so trust is decided on lineage the agent does not control
(Table~\ref{tab:substitution}). A clean context, with no decode origin, still allows
a legitimate critical call. This is the concrete difference between verifying a
label and maintaining lineage---the distinction on which the central guarantee
rests.

\par\addvspace{7pt}\noindent\begin{minipage}{\columnwidth}\centering
\captionof{table}{Valid-label substitution: enforcement that trusts the
agent-supplied binding vs.\ layer-maintained lineage.}
\label{tab:substitution}
\footnotesize
\begin{tabular}{@{}p{0.66\columnwidth}l@{}}
\toprule
Enforcement & Decision \\
\midrule
Agent-supplied binding only (naive)        & \textbf{allow (leaks)} \\
Layer-maintained, dependency omitted       & block \\
Layer-maintained, dependency in arguments  & block \\
Layer-maintained, clean context            & allow \\
\bottomrule
\end{tabular}
\end{minipage}\par\addvspace{7pt}

\subsection{Component ablation}
The mechanism has four components---decode marking, cryptographic authentication
(the HMAC), trusted lineage tracking (the layer's ledger and context taint), and
sink mediation---and each defends a distinct attack. Table~\ref{tab:ablation}
disables one component at a time in the full implementation and measures three
attacks: the honest in-domain attack, valid-label substitution (in-domain), and a
forged \textsc{Trusted} label carried across an untrusted relay to a remote verifier
that holds only the presented binding (cross-domain). The full mechanism blocks all
three. Removing marking leaks everything (nothing is flagged untrusted). Removing
mediation leaks everything (no enforcement point). The two middle rows are the ones
that disambiguate the design: removing the HMAC still blocks the in-domain attacks,
because the ledger decides, but leaks the \emph{cross-domain forged relay}, because
a remote verifier without the ledger has only the signature to trust; removing
lineage tracking (reverting to deciding on the agent-presented binding) still blocks
forgery via the HMAC but leaks \emph{valid-label substitution}, because a genuine
\textsc{Trusted} binding from an unrelated tool passes an HMAC check. Cryptographic
authentication and layer-maintained lineage are therefore both necessary and
non-interchangeable: the HMAC alone does not stop substitution, and the ledger alone
does not survive a cross-domain relay. The taint baseline of Table~\ref{tab:asr}
corresponds to the ``$-$ cryptographic auth'' row.

\par\addvspace{7pt}\noindent\begin{minipage}{\columnwidth}\centering
\captionof{table}{Four-component ablation: effect of disabling each component on three attacks (blocked vs.\ LEAK).}
\label{tab:ablation}
\footnotesize
\begin{tabular}{@{}p{0.34\columnwidth}ccc@{}}
\toprule
Configuration & Honest & Substitution & Forged relay \\
\midrule
\textbf{Full mechanism}           & blocked & blocked & blocked \\
$-$ decode marking                & LEAK    & LEAK    & LEAK \\
$-$ cryptographic auth (HMAC)     & blocked & blocked & \textbf{LEAK} \\
$-$ trusted lineage tracking      & blocked & \textbf{LEAK} & blocked \\
$-$ sink mediation                & LEAK    & LEAK    & LEAK \\
\bottomrule
\end{tabular}
\end{minipage}\par\addvspace{7pt}

\subsection{Cross-context influence propagation}
A natural question is whether taint survives an inter-agent message boundary, not
just a tool-return handle. It does: Table~\ref{tab:crosscontext} shows Agent~A
decoding an untrusted instruction and handing it to Agent~B in a separate execution
context. When the message is carried on a \emph{mediated} channel---through the
layer, as assumptions~(iii) and~(iv) require of inter-agent communication---the
layer binds it to its source and propagates the taint, so B's critical call is
blocked (\textsc{untrusted context lineage}). This is the result: context taint
crosses every mediated boundary, agent-to-agent messages included. The second row
marks the edge of the model rather than a result: a channel the layer does not
mediate (A paraphrasing into free text on a side path) is outside assumptions~(iii)
and~(iv) by construction---in a fully mediated deployment no such channel exists, and
a deployment that leaves one unmediated has simply not placed the control on it.
Cryptographic authenticity cannot stand in for a dependency that was never recorded,
so the scope is stated plainly with the invariant: the guarantee is for fully
mediated execution contexts. Extending capture into in-model paraphrase is discussed
in Section~\ref{sec:limits}.

\par\addvspace{7pt}\noindent\begin{minipage}{\columnwidth}\centering
\captionof{table}{Cross-context propagation: an untrusted decode in Agent~A reaches a
critical call in Agent~B's context.}
\label{tab:crosscontext}
\footnotesize
\begin{tabular}{@{}p{0.62\columnwidth}l@{}}
\toprule
Channel from A to B & B's critical call \\
\midrule
Mediated inter-agent message (in scope)       & block (taint propagated) \\
Unmediated side channel (outside the model)   & not covered \\
\bottomrule
\end{tabular}
\end{minipage}\par\addvspace{7pt}

\subsection{State transitions}
Table~\ref{tab:states} walks one benign-worded payload through three system
states. Pre-attack, the clean logo yields a normal summary and no critical call.
Post-attack undefended, the payload decodes and the critical sink executes,
leaking secrets. Post-attack defended, the payload decodes identically but the
verifier blocks the sink; no secrets leave. The decoded instruction is present in
the agent's context in both post-attack rows---the defense does not try to keep it
out---so the only difference between a breach and a safe run is whether the
critical call is permitted to fire, which is precisely the boundary the binding
governs.

\par\addvspace{7pt}\noindent\begin{minipage}{\columnwidth}\centering
\captionof{table}{System state across clean, undefended, and defended runs.}
\label{tab:states}
\footnotesize
\begin{tabular}{lll}
\toprule
State & Critical call & Secrets leaked \\
\midrule
Pre-attack (clean logo)      & none                     & no  \\
Post-attack (undefended)     & \texttt{export\_secrets} executed & \textbf{yes} \\
Post-attack (defended)       & blocked                  & no  \\
\bottomrule
\end{tabular}
\end{minipage}\par\addvspace{7pt}

\subsection{Tamper-resistance across handoffs}
Table~\ref{tab:tamper} evaluates a malicious relaying agent. A genuine
\textsc{Untrusted} binding is blocked at a critical sink; an attempt to forge a
\textsc{Trusted} label or to strip the HMAC fails verification and is also
blocked; a genuinely trusted origin (a non-decode tool such as authenticated user
input) is allowed. The binding therefore survives relaying intact.

\par\addvspace{7pt}\noindent\begin{minipage}{\columnwidth}\centering
\captionof{table}{Verifier decision under relay tampering.}
\label{tab:tamper}
\footnotesize
\begin{tabular}{ll}
\toprule
Relay action & Verifier decision \\
\midrule
Genuine \textsc{Untrusted} label      & block \\
Forged \textsc{Trusted} label         & block (HMAC fails) \\
Stripped HMAC                         & block (HMAC fails) \\
Honest \textsc{Trusted} origin        & allow \\
\bottomrule
\end{tabular}
\end{minipage}\par\addvspace{7pt}

\subsection{Carrier distortion}
Table~\ref{tab:imp} reports peak signal-to-noise ratio (PSNR) and structural
similarity (SSIM) between cover and stego logos. The metadata carrier is lossless
(the pixels are unchanged); the LSB carrier is near-lossless ($\approx$76\,dB); the
DCT carrier trades a lower PSNR ($\approx$18\,dB, SSIM $\approx$0.94) for robustness
to the pixel round-trip. We stress that PSNR and SSIM bound \emph{pixel-level
distortion} and do not by themselves establish perceptual imperceptibility,
particularly for the DCT carrier at $\approx$18\,dB; we therefore treat these as
distortion metrics, offer the amplified pixel-difference visualization
(Fig.~\ref{fig:coverstego}) as the perceptual check, and leave a formal subjective
study to future work. The security guarantee does not depend on imperceptibility:
containment blocks the payload whether or not a human would notice the carrier.

\par\addvspace{7pt}\noindent\begin{minipage}{\columnwidth}\centering
\captionof{table}{Carrier distortion between cover and stego logos (distortion
metrics, not a perceptual claim).}
\label{tab:imp}
\footnotesize
\begin{tabular}{lcc}
\toprule
Carrier & PSNR (dB) & SSIM \\
\midrule
LSB       & high ($\approx$76)  & $\approx$1.000 \\
DCT       & moderate ($\approx$18) & $\approx$0.94 \\
Metadata  & lossless (pixels unchanged) & 1.000 \\
\bottomrule
\end{tabular}
\end{minipage}\par\addvspace{7pt}

A defender nonetheless cannot rely on visual inspection or a perceptual-hash gate
to catch the metadata and LSB carriers, which sit at or near the lossless ceiling;
the DCT carrier is the one a perceptual gate might flag, which is why we do not
claim it is imperceptible. Fig.~\ref{fig:imp} renders the per-carrier PSNR and SSIM,
with the lossless metadata bar capped so the axis stays readable.

\par\addvspace{7pt}\noindent\begin{minipage}{\columnwidth}\centering
\includegraphics[width=0.9\columnwidth]{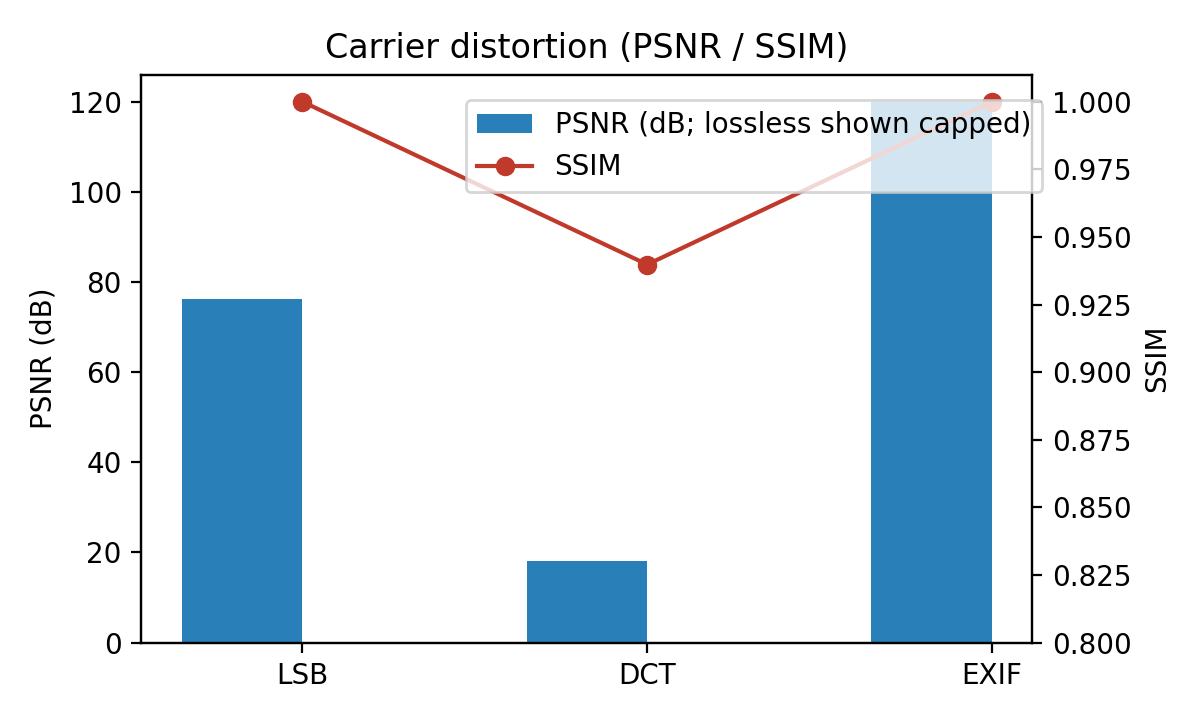}
\captionof{figure}{Per-carrier distortion (PSNR and SSIM).}
\label{fig:imp}
\end{minipage}\par\addvspace{7pt}

\subsection{Overhead}
Table~\ref{tab:overhead} reports the cost of the mechanism, measured on commodity
hardware over $2\times10^4$ iterations after warm-up. Minting a binding and
verifying it are pure HMAC-SHA256 operations over a short canonical string, each
under $10\,\mu$s; their combined cost of $\approx\!17\,\mu$s is added once per
critical call. For reference, a single decode tool call over the MCP stdio
transport (warm session) takes a median of $\approx\!11$\,ms, so the binding adds
about $0.16\%$
to a tool invocation and is negligible beside model inference, which is orders of
magnitude larger. These figures are for the \emph{cryptographic operations}; the
full enforcement path, which also inserts into the ledger and traverses the
dependency closure, is characterized separately below and is what a deployment
should budget for.

\par\addvspace{7pt}\noindent\begin{minipage}{\columnwidth}\centering
\captionof{table}{Mechanism overhead per operation.}
\label{tab:overhead}
\footnotesize
\begin{tabular}{lr}
\toprule
Operation & Cost \\
\midrule
\textsc{MintBinding} (HMAC sign)         & $\approx 8.3\,\mu$s \\
\textsc{EnforceCritical} (HMAC verify)   & $\approx 8.7\,\mu$s \\
Combined per critical call               & $\approx 17\,\mu$s \\
MCP decode round-trip (warm)             & $\approx 11$\,ms \\
Added overhead vs.\ one decode call      & $\approx 0.16\%$ \\
\bottomrule
\end{tabular}
\end{minipage}\par\addvspace{7pt}

The revised mechanism does more than sign: \textsc{MintBinding} inserts into the
ledger and \textsc{EnforceCritical} walks the dependency closure and checks context
taint. We therefore benchmark the complete path as these grow
(Table~\ref{tab:scaling}). Minting stays near $11\,\mu$s regardless of chain depth
(ledger insertion is amortized $O(1)$). Enforcement is linear in the size of the
dependency closure it traverses: $\approx\!10\,\mu$s at depth~1, rising to
$\approx\!0.68$\,ms at depth~100 and $\approx\!6.6$\,ms at depth~1000---the last
comparable to a single decode round-trip and reached only by pathologically deep
lineage. It is effectively flat ($\approx\!6\,\mu$s) in the number of concurrent
tainted contexts, since the taint lookup is keyed by context. For realistic lineages
(closure depth in the tens) the full path stays well under $0.1$\,ms.

\par\addvspace{7pt}\noindent\begin{minipage}{\columnwidth}\centering
\captionof{table}{Full enforcement-path cost vs.\ dependency-closure depth and
number of concurrent tainted contexts.}
\label{tab:scaling}
\footnotesize
\begin{tabular}{lrrrr}
\toprule
 & \multicolumn{4}{c}{closure depth} \\
\cmidrule(l){2-5}
Operation & 1 & 10 & 100 & 1000 \\
\midrule
\textsc{MintBinding} ($\mu$s)      & 10.7 & 12.0 & 12.9 & 10.9 \\
\textsc{EnforceCritical} ($\mu$s)  & 9.5  & 72.6 & 680  & 6605 \\
\midrule
 & 1 & 100 & 1000 & \\
\cmidrule(l){2-4}
\textsc{EnforceCritical} vs.\ contexts ($\mu$s) & 6.2 & 6.4 & 6.3 & \\
\bottomrule
\end{tabular}
\end{minipage}\par\addvspace{7pt}

\subsection{Decode-set completeness}
The security invariant assumes every concealment-reversing tool is registered in
the decode set $\mathcal{D}$. Because this is the assumption a reviewer will press,
we exercise it directly. Table~\ref{tab:decodeset} reports three configurations
against the same stego attack and a legitimate critical call. With the correct
$\mathcal{D}$ the attack is blocked and the legitimate call is allowed. If the
attacker reaches the payload through a decoder that is \emph{missing} from
$\mathcal{D}$ (e.g.\ an unregistered audio or document extractor), its output is
minted \textsc{Trusted} and the attack \emph{leaks}---the failure is an
unregistered decoder, not a defeat of the mechanism. Conversely, an
\emph{over-broad} $\mathcal{D}$ that labels a benign non-decode tool
\textsc{Untrusted} falsely blocks a legitimate critical call. The set is thus the
single knob that trades missed-decoder risk against false blocks, which is why we
recommend it be a small, audited allow-list fixed at tool registration.

\par\addvspace{7pt}\noindent\begin{minipage}{\columnwidth}\centering
\captionof{table}{Decode-set completeness: outcome vs.\ configuration.}
\label{tab:decodeset}
\footnotesize
\begin{tabular}{lcc}
\toprule
Decode set $\mathcal{D}$ & Stego attack & Legitimate critical call \\
\midrule
Correct                              & blocked          & allowed \\
Missing decoder (unregistered)       & \textbf{leaks}   & allowed \\
Over-broad (benign tool included)    & blocked          & \textbf{false block} \\
\bottomrule
\end{tabular}
\end{minipage}\par\addvspace{7pt}

\subsection{Label integrity across handoffs}
We next stress the binding under relaying. Table~\ref{tab:integrity} shows the
verifier decision after honest relays and under a compromised worker. A genuine
binding verifies and is correctly blocked after 1, 3, 5, and 10 honest hops---the
label travels intact. A compromised worker that rewrites the decoded text,
promotes the label to \textsc{Trusted}, or strips the binding entirely always
fails verification and is blocked, because none of these can reproduce the
server-held HMAC. This complements Table~\ref{tab:tamper} by showing integrity is
preserved across depth, not just at a single hop.

\par\addvspace{7pt}\noindent\begin{minipage}{\columnwidth}\centering
\captionof{table}{Label integrity across relay hops and under a compromised worker.}
\label{tab:integrity}
\footnotesize
\begin{tabular}{ll}
\toprule
Relay scenario & Verifier decision \\
\midrule
Honest relay, 1 / 3 / 5 / 10 hops          & verifies; critical call blocked \\
Compromised: rewrite payload, keep HMAC    & block (HMAC mismatch) \\
Compromised: forge \textsc{Trusted} label  & block (HMAC mismatch) \\
Compromised: strip binding                 & block (no valid label) \\
\bottomrule
\end{tabular}
\end{minipage}\par\addvspace{7pt}

\subsection{Legitimate-task utility}
A containment control is only deployable if it does not block legitimate work.
Table~\ref{tab:legit} reports three legitimate critical tasks, one of which---an
authorized payment that uses an optical-character-recognition (OCR)'d invoice
number---legitimately derives from a decode tool. Without the trusted-promotion
path, that task is falsely blocked (67\% completion, one false block), the expected
cost of treating every decode origin as untrusted. With an audited promotion that
re-labels the specific OCR'd value \textsc{Trusted}, all three tasks complete
(100\%, zero false blocks) while an unauthorized attack from the same decode origin
remains blocked, because the attacker cannot invoke the promotion. Promotion is thus
the deployable complement to containment: it restores utility at a controlled,
logged point without reopening the vector.

\par\addvspace{7pt}\noindent\begin{minipage}{\columnwidth}\centering
\captionof{table}{Legitimate-task completion and false blocks, with and without the
audited trusted-promotion path.}
\label{tab:legit}
\footnotesize
\begin{tabular}{lccc}
\toprule
Regime & Completion & False blocks & Attack blocked \\
\midrule
No promotion              & 2/3 (67\%)  & 1 & yes \\
\textbf{Trusted promotion} & \textbf{3/3 (100\%)} & \textbf{0} & yes \\
\bottomrule
\end{tabular}
\end{minipage}\par\addvspace{7pt}

\subsection{Realistic decode trigger}
Section~\ref{sec:threat} assumes a worker invokes a decode step on a signature
logo; a reviewer may reasonably ask why an ordinary summarizer would do so. We
separate realistic preconditions from the controlled proof-of-concept. The
metadata, Base64, and OCR carriers need no bespoke decoder: a summarizer that
extracts text from an image (OCR) or parses embedded metadata to caption or
describe an attachment surfaces these payloads as a side effect of normal
multimodal handling, and contemporary agents routinely route images to exactly such
tools. We model this in the testbed with a content-type router that dispatches an
inline image to an image-analysis tool, which in turn invokes the registered
extractor; the decoded text then enters the agent's context. The LSB and DCT
carriers are stronger-attacker variants that require a decoder matching the
concealment scheme; we include them to stress the mechanism across the spatial and
frequency branches, not as claims about default summarizer behavior. In every case
the defense acts identically, because registration in $\mathcal{D}$---not the reason
the tool ran---determines the label.

\section{Deployment}
\label{sec:deployment}
The mechanism is a drop-in addition to an MCP server and its tool-call boundary;
it requires no change to the agent model or the orchestration logic.

\noindent\textbf{Where it lives.} \textsc{MintBinding} runs inside each tool
handler at return time and writes to the layer's ledger; \textsc{EnforceCritical}
runs in the client's tool-dispatch path immediately before a call executes and
reads that ledger. Both are a few lines over a standard HMAC library plus an
in-memory ledger keyed by execution context. The signing key $k$ is generated once
per server process and never leaves it; bindings travel with tool results as an
extra signed field, transparent to agents that do not inspect it, while the ledger
and context taint stay inside the layer and are never exposed to agents.

\noindent\textbf{Configuring $\mathcal{D}$ and $\mathcal{C}$.} The decode set
$\mathcal{D}$ enumerates tools that reverse a concealment transform (stego and
metadata extractors, Base64 decoders, OCR); the critical set $\mathcal{C}$
enumerates side-effecting sinks (secret export, outbound email, file write,
payment). Both are short, server-authored allow/deny lists reviewed at tool
registration---the same point at which a tool's schema is already vetted---so no
runtime classification is needed.

\noindent\textbf{Trusted promotion.} When a decoded artifact must legitimately
drive an action (e.g.\ an OCR'd invoice number the operator asked to pay), an
explicit, audited promotion step re-labels that specific value \textsc{Trusted}
under human or policy authorization and records the clearance in the context
ledger. The promotion is logged with the original binding, so the provenance trail
is preserved. We implement and evaluate this path rather than leaving it as a
proposal: it raises legitimate-task completion to 100\% with no loss of protection
(Table~\ref{tab:legit}).

\noindent\textbf{Distributed enforcement.} When tools span multiple hosts,
verification uses either a shared key over a secured channel or a public-key
signature in place of the HMAC; the algorithms are unchanged apart from the
sign/verify primitive.

\section{Design Trade-offs}
\label{sec:tradeoffs}

\noindent\textbf{Containment vs.\ detection.} Binding contains rather than
detects, so it is robust to unseen carriers and wording, but it cannot recover
the \emph{benign} information a decoded artifact might also carry: a legitimately
hidden watermark read by a decode tool is labeled \textsc{Untrusted} like any
other. Deployments that need such content must add an explicit, audited trusted-promotion
path.

\noindent\textbf{Scope of protection.} The mechanism guards \emph{actions}, not
\emph{summary text}. A laundered instruction can still corrupt the natural-language
summary the agent returns; binding prevents it from reaching a critical tool, not
from influencing prose. This bounds the defense to the excessive-agency and
disclosure harms, which is the right scope for a tool-boundary control.

\noindent\textbf{Decode-set completeness.} Security rests on every
concealment-reversing tool being registered in $\mathcal{D}$. An unregistered
decoder is a gap; conversely, over-broad registration (labeling non-decoding
tools \textsc{Untrusted}) raises false blocks. The set is small and auditable in
practice, but its maintenance is an operational cost.

\noindent\textbf{Key management.} Tamper-evidence depends on confining the HMAC
key to the MCP layer. Distributing enforcement across hosts requires a shared-key
or signature-verification deployment, trading simplicity for scalability.

\noindent\textbf{Lineage capture.} Soundness against valid-label substitution
rests on the layer maintaining the complete lineage of each invocation
(assumption~iv): the resolved argument dependencies and the decode origins entering
each context. This requires the layer to mediate tool I/O and keep a per-context
ledger---more than a stateless signature check, though still far cheaper than
inference. Its residual is influence the layer cannot observe: if decoded text
leaves the mediated path entirely (e.g.\ an agent paraphrases it out of band into a
context the ledger does not cover), context taint cannot record it. We therefore
treat context taint as a conservative over-approximation that errs toward blocking,
and note that completeness of lineage capture, not the cryptography, is the property
a deployment must get right.

\noindent\textbf{Granularity.} We bind at message/tool-result granularity, not at
token granularity inside model reasoning. This is simple and model-agnostic but
cannot express partial trust within a single decoded blob.

\section{Limitations}
\label{sec:limits}
Our evaluation remains a controlled testbed: 24 payloads across three carriers and
eight evasions, one cover logo, one small local model, and a keyword/regex scanner
and an in-process taint tracker as baselines. This is broader than a single attack
bank---it adds a label-substitution stress test, a four-component ablation, a
legitimate-task evaluation, and an information-flow baseline---but it is still not a
production-scale measurement: larger and more varied payload banks, multiple cover
images and model families, a modern learned or provenance-aware detector as a
baseline, and---most important for practical impact---a demonstration inside an
existing production email-agent workflow rather than our controlled summarizer, all
remain future work, and operational-stealth measurement
frameworks~\cite{stealthbench2026} suggest one axis for that expansion. Our proof sketch reduces security to two properties: the
infeasibility of forging the HMAC under key confinement, and the completeness of
lineage capture by the layer (assumption~iv). The former is standard; the latter is
an implementation obligation, and we are explicit that the guarantee is scoped to
fully mediated execution contexts. Within that scope, taint propagates across every
mediated boundary, including inter-agent messages (Table~\ref{tab:crosscontext}). A
channel the layer does not mediate is outside the model by construction; the one
substantive extension it points to is \emph{in-model} influence---an agent
paraphrasing a decoded instruction into its own output---which a mediated side
channel would not capture and which tainting an agent's output whenever its input
context is tainted would address. We leave that, a fully machine-checked proof, and
a formal treatment of lineage completeness to future work. Finally, the
defense protects \emph{actions}, not \emph{summary integrity}: a laundered
instruction it blocks at the sink can still skew the natural-language summary the
agent returns. We bound this harm (no critical action executes) but do not
eliminate it; quantifying the residual summary-corruption rate requires running the
agent model and is a measurement the released notebook supports and that we leave to
future work.

\section{Conclusion}
\label{sec:conclusion}
Provenance defenses for LLM agents have assumed that untrusted content arrives
through a text channel, but a hidden instruction can ride into a multi-agent MCP
pipeline inside a benign-looking email-signature logo, pass text content scanners
that see only pixels, and become instructions only when a tool decodes it
— at which point, in an undefended pipeline, it is laundered into the trusted control
flow and drives a critical tool call; we characterized this cross-modal
trust-laundering vector, proposed decode-time provenance binding in which the MCP
layer mints a cryptographically signed untrusted label at the concealment-reversal
boundary, and independently maintains the provenance lineage of each critical call,
enforcing at the critical tool boundary that no call with untrusted lineage
executes. We showed on a real MCP testbed with three carriers and 24 adaptive
payloads that the mechanism reduces attack success from 100\% (undefended) and
62.5\% (keyword scanner and regex parser) to 0\%, resists valid-label substitution
by a compromised agent because trust is decided on layer-held lineage rather than
an agent-presented label, survives label forgery and stripping across agent
handoffs, and — via an audited trusted-promotion path — preserves 100\% of
legitimate-task completion; together these demonstrate that, within a fully mediated
MCP execution environment, containment at the decode boundary backed by
layer-maintained lineage — rather than detection of the carrier — is the property
that holds, with all code and data released for reproduction.

\clearpage
\bibliographystyle{IEEEtran}
\bibliography{references}

\end{document}